\documentclass[printer]{aa}
\usepackage[fleqn]{amsmath}
\usepackage{amssymb}
\usepackage[none]{hyphenat}
\usepackage{times}
\usepackage{graphicx}
\usepackage{color}
\usepackage{blindtext}
\usepackage{textcomp}
\usepackage{gensymb}
\usepackage[varg]{txfonts}
\usepackage{natbib}
\bibpunct{(}{)}{;}{a}{}{,} 

\newcommand{\revised}{\color{black}}

\begin{document}

\title{Understanding the trans-Neptunian Solar system}
\subtitle{Reconciling the results of serendipitous stellar occultations and the inferences from the cratering record}
\titlerunning{Reconciling occultations and craters}
\author{Andrew Shannon, Alain Doressoundiram, Fran\c{c}oise Roques, Bruno Sicardy, Quentin Kral}
\authorrunning{Shannon et al. }


\institute{LESIA, Observatoire de Paris, Universit\'{e} PSL, CNRS, Sorbonne Universit\'{e}, Universit\'{e} de Paris, 5 place Jules Janssen, 92195 Meudon, France}

\date{31 December 2020 / ? ? ?}

\abstract{The most pristine remnants of the Solar system's planet formation epoch orbit the Sun beyond Neptune, the small bodies of the trans-Neptunian object populations.  The bulk of the mass is in $\sim 100~\rm{km}$~objects, but objects at smaller sizes have undergone minimal collisional processing, with ``New Horizons'' recently revealing that $\sim 20~\rm{km}$~{\revised effective diameter body} (486958) Arrokoth appears to be a primordial body, not a collisional fragment.  This indicates bodies at these sizes (and perhaps smaller) retain a record of how they were formed, and are the most numerous record of that epoch.  However, such bodies are impractical to find by optical surveys due to their very low brightnesses.  Their presence can be inferred from the observed cratering record of Pluto and Charon, and directly measured by serendipitous stellar occultations.  These two methods produce conflicting results, with occultations measuring roughly ten times the number of $\sim\rm{km}$~bodies inferred from the cratering record.  We use numerical models to explore how these observations can be reconciled with evolutionary models of the outer Solar system.  We find that models where the initial size of bodies decreases with increasing semimajor axis of formation, and models where the surface density of bodies increases beyond the 2:1 mean-motion resonance with Neptune can produce both sets of observations, though comparison to various observational tests favours the former mechanism.  We discuss how to evaluate the astrophysical plausibility of these solutions, and conclude extended serendipitous occultation surveys with broad sky coverage are the most practical approach.}


   \maketitle

\section{Introduction}

The Solar system beyond Neptune is thought to contain the least collisionally evolved population of small bodies in the Solar system.  Owing to their relatively unevolved state, observations of these bodies thus give the best direct constraints of how planetesimals formed and evolved at sizes of $\sim 1 \sim 100~\rm{km}$.  There is an observed turnover in the size-number distribution around a radius of $s = 50 \sim 70~\rm{km}$ \citep{2004AJ....128.1364B,2014ApJ...782..100F}.  Although it was originally suspected this could be of collisional origin, further work has shown it is far more likely to be of primordial origin \citep{2012A&A...540A..30V}, as it is at too large a size to be produced in even the most optimistic models of collisional evolution \citep{2005Icar..173..342P}.  Similar conclusions have been reached about the size-number distribution of the  asteroids \citep{2009Icar..204..558M}.  The conclusion that the turnover in the size-number distribution of trans-Neptunian objects is not collisional is strongly re-enforced  by \textit{New Horizons}~observations of the $\sim 20~\rm{km}$~body (486958) Arrokoth, which appears to be a primordial body rather than a collisional fragment \citep{2019Sci...364.9771S}. This may also be true of the $\sim 4~\rm{km}$~Jupiter family comet\footnote{with probable origin in the trans-Neptunian belt \citep{1997Sci...276.1670D}} 67P/Churyumov-Gerasimenko \citep{2015Natur.526..402M}, although that interpretation has been contested \citep{2017A&A...597A..62J}.  \citet{2021Icar..35614256M} estimate from the survival of (486958) Arrokoth that the size where bodies transition from being mostly primordial to mostly collisional fragments may be $\sim 1~\rm{km}$.  Further evidence of the low amount of collisional evolution of trans-Neptunian objects comes from the survival of ultra-wide binaries in the Cold Classical Kuiper belt \citep{2012ApJ...744..139P}.  

Amongst the populations of trans-Neptunian objects, the evidence of minimal collisional evolution applies most clearly to the Cold Classical Kuiper belt.  In the usual population classification scheme \citep{2008ssbn.book...43G}, the Cold Classicals are known to exist mainly within the Main Classical Kuiper belt, between the 3:2 and 2:1 Mean Motion Resonances with Neptune.  A few appear to exist beyond the 2:1 \citep{2018ApJS..236...18B}, but there the properties of the belt are less clear, and \citet{nesvorny2020eccentric} suggested they may have been implanted from primordial orbits closer to the Sun.  The Cold Classicals formed close to their current orbits \citep{2012ApJ...750...43D, 2015AJ....150...68N}, isolating them from much of the Solar system's evolution.  The various hot populations (Resonant, Hot Classicals, Scattering, Detached) formed closer to the Sun before being pushed to their current orbits \citep{1993Natur.365..819M, 2008Icar..196..258L,2014AJ....148...56B,2019AJ....158...64V}, and could have undergone significant early collisional evolution \citep{2007Icar..188..468C,2012MNRAS.423.1254C,2019MNRAS.485.5511S}, depending on the timing and details of the outer Solar system's evolution \citep{2005Natur.435..466G,2018Icar..305..262M,2018NatAs...2..878N}.  The observational evidence on the collisional evolution of the Hot population, is rather unclear.  The hot populations lack ultra-wide binaries \citep{2019Icar..334...62G}, but these may have been lost by a non-collisional process \citep{2010ApJ...722L.204P}.  A few of the largest bodies show evidence of giant impacts \citep{2005Sci...307..546C, 2007Natur.446..294B}, but it appears these may have occurred at low speeds \citep{2011AJ....141...35C,2019AJ....157..230P,2020NatAs...4...89P}, and thus reflect accretional collisions rather than destructive collisions, although this question remains far from settled.

Thus, well constrained measurements of the size-number distribution of trans-Neptunian objects, especially Cold Classicals, at sizes well below $\sim 50~\rm{km}$~could provide strong support for models of their formation; planetesimal formation via the streaming instability \citep{2005ApJ...620..459Y, 2007Natur.448.1022J} is currently generally seen as the most promising mechanism for forming planetesimals at these sizes \citep{2022ASSL..466....3R}, and makes clear predictions for the expected size-number distribution {\revised below the primordial break in the size-number distribution} \citep{2016ApJ...822...55S,2017ApJ...847L..12S}.  Conversely, of course, such measurements might instead provide evidence of a competing mechanism, such as fluffy aggregate growth \citep{2016ApJ...832L..19A}, collisional pebble agglomeration \citep{2016ApJ...818..175S}, or turbulent clustering \citep{2010Icar..208..518C,2016LPI....47.2661C,2020ApJ...892..120H}.

At sizes below tens of kilometres, however, it is prohibitively difficult to strongly constrain size-number distribution with reflected light surveys \citep{2004AJ....128.1364B, 2014ApJ...782..100F}.  Two techniques can be employed to measure the size-number distribution down to sizes of a couple hundred meters.  Small TNOs can be detected by serendipitous occultations of stars \citep{2003ApJ...594L..63R, 2009Natur.462..895S}, and counting the size-number distribution of craters on Pluto and Charon seen by \textit{New Horizons} to infer the size-number distribution of the impactors \citep{2017Icar..287..187R}.  So far, four experiments have reported detections of serendipitous occultations \citep[][]{2012ApJ...761..150S, 2015MNRAS.446..932L, 2019NatAs...3..301A, 2020submitted}, and at least one more under construction \citep{2019EPSC...13.1089L}, while others provided upper limits \citep{2008AJ....135.1039B,2009AJ....138..568B,2010AJ....139.2003W,2013MNRAS.429.1626C,2013AJ....146...14Z}.

\begin{figure}
    \centering
    \includegraphics[width=\linewidth]{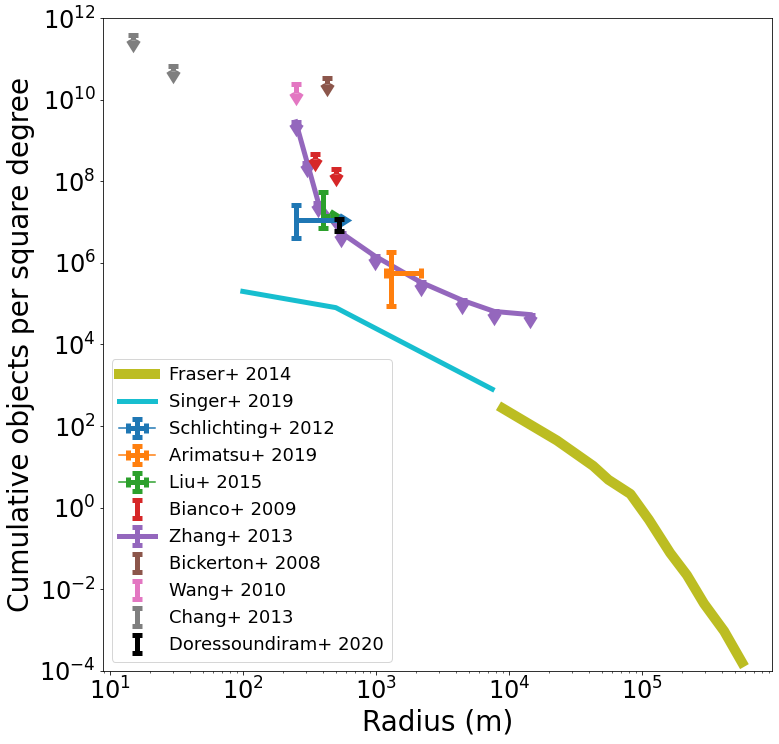}
    \caption{Existing measurements of the size-number distribution of small bodies in the outer Solar system, comparing the results of occultation surveys (upper limits: \citealt{2008AJ....135.1039B,2009AJ....138..568B,2010AJ....139.2003W,2013MNRAS.429.1626C,2013AJ....146...14Z}; and detections: \citealt{2012ApJ...761..150S, 2015MNRAS.446..932L,2019NatAs...3..301A,2020submitted}), optical surveys \citep{2014ApJ...782..100F}, and crater counting \citep{2019Sci...363..955S} which is converted to a population number using the dynamical model of \citet{2015Icar..258..267G}.  Because the different types of surveys use different measurements, they can only be directly compared in a somewhat model dependent way.  Thus, the apparent disagreement may represent a disagreement of the observations, or an incorrect assumption in the modelling.}
    \label{fig:smallBodyObservations}
\end{figure}

The serendipitous occultation surveys have found generally results compatible with one another, which give the number of $\sim\rm{km}$~objects to be $10 \sim 100 \times$~as numerous as inferred from the cratering record \citep{2019Sci...363..955S} (Figure \ref{fig:smallBodyObservations}).  Reconciling these two sets of observations may be possible as they do not necessarily sample the same populations of bodies{\revised, as the occultations measure the present day population along the line of sight, and the craters are a time-integrated record of bodies overlapping the orbit of Pluto-Charon}.  Here we investigate whether a hypothetical extension of the Cold populations to larger semimajor axes can be consistent with both sets of observations, allowing for the possibility that the characteristic formation size of bodies might decrease (or increase) with their semimajor axis of formation.  In \textsection \ref{sec:methods}, we outline the method we use to model the evolution of the trans-Neptunian populations and compare them to observations.  In \textsection \ref{sec:results}, we present the results of the modelling.  In \textsection \ref{sec:discussion}, we discuss the implications of the results and how they can be validated.  In \textsection \ref{sec:conclusions}, we present our conclusions.

\section{Methods}

\label{sec:methods}

\subsection{Single Model Evaluation}

To simulate the collisional and dynamical evolution of the outer Solar system, we use a statistical particle-in-a-box approach where bodies of similar masses and orbital elements are grouped together \citep{1969edo..book.....S}.  Our code is derived from the code presented in \citet{2015ApJ...801...15S, 2016ApJ...818..175S} with several modifications to make it suitable for the problem at hand.

While \citet{2015ApJ...801...15S, 2016ApJ...818..175S} considered only a single radial zone, here we consider four unique, mutually interacting, dynamical populations: the Cold Classical Kuiper belt, the Hot Classical Kuiper belt, the Scattered disk, and the Resonant disk.  Within each population, there are multiple semimajor axis bins to represent bodies at different distances from the Sun; most of the bin position choices do not significantly affect the evolution, they are set to divide the modelled population in a uniform way while keeping the total number of bins small enough to keep computations tractable.  The  inner-most bin of the Cold Classical Kuiper belt is set to have an outer end near 50 au to accommodate our supposed possibility of a discontinuity in the surface density of Cold Classicals there.  Given that, we set the Cold Classical to have 6 bins at $42~\rm{au} < a < 52~\rm{au}$, $52~\rm{au} < a < 62~\rm{au}$, $62~\rm{au} < a < 72~\rm{au}$, $72~\rm{au} < a < 82~\rm{au}$, $82~\rm{au} < a < 92~\rm{au}$, $92~\rm{au} < a < 102~\rm{au}$.  The first bin covers the known Cold Classicals, the subsequent bins cover our hypothetical extention.  The Hot Classicals have 2 bins for objects with $36~\rm{au} < a < 48~\rm{au}$, and $48~\rm{au} < a < 60~\rm{au}$.  For the Scattered disk we use 2 bins, one for objects with $35~\rm{au} < a < 62.5~\rm{au}$, and one for $62.5~\rm{au} < a < 100~\rm{au}$; although there are Scattered objects with larger semimajor axes, the low surface density and long orbital periods renders them relatively unimportant for any collisional evolution.  The Resonant disk has 3 bins to represent the populations of the 3:2 MMR from $39~\rm{au}$~to $40~\rm{au}$, 2:1 from $47.3~\rm{au}$~to $48.3~\rm{au}$, and the 5:2 from $54.9~\rm{au}$~to $55.9~\rm{au}$.  For the resonant populations the eccentricity rather than semimajor axis bin width sets the radial extent of the population, so the choice of bin width is largely unimportant.  Each semimajor axis bin is assigned a single eccentricity and inclination; neither of which evolves.  All Cold Classical bins were assigned $e = 0.04$~and $i = 2.5 \degree$ \citep{2011AJ....142..131P}, all Hot Classical bins were assigned $e = 0.2$~and $i = 14.5 \degree$ \citep{2017AJ....153..236P}, Scattered bins were assigned $i = 14.5 \degree$~and eccentricities chosen so the pericentre was $q = 30~\rm{au}$, and for the resonant objects, the 3:2s were assigned $e = 0.175$~and $i = 12 \degree$, the 2:1s were assigned $e = 0.28$~and $i = 4.0 \degree$, and the 5:2s were assigned $e = 0.35$~and $i = 10\degree$~\citep{2016AJ....152...23V}.

\begin{figure}
  \includegraphics[width=\linewidth]{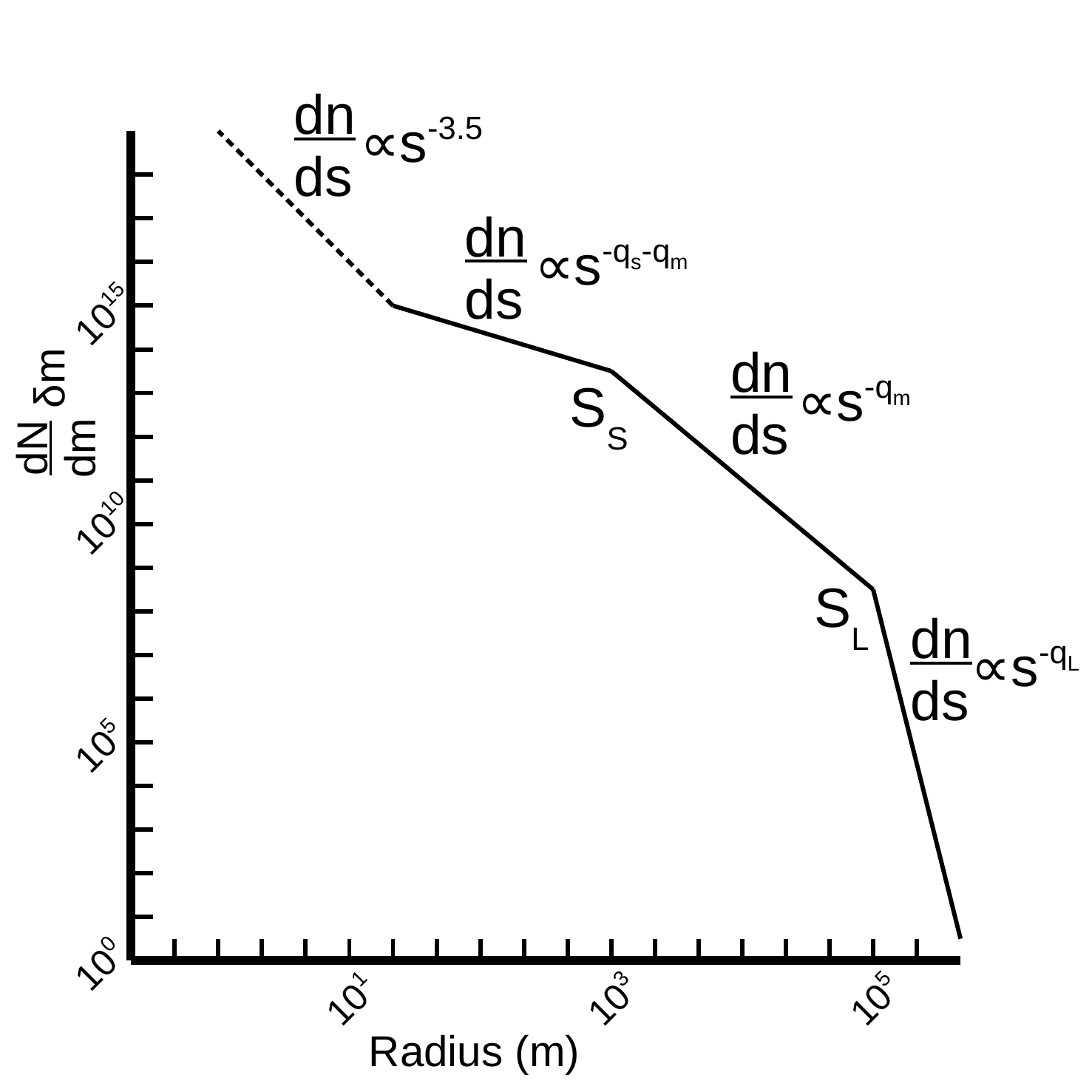}
  \caption{An example sketch of how we parameterise the initial size-number distribution in our simulations.  Bodies in the solid line region are placed in bins logarithmically with six bins per decade of mass.  {\revised The intermediate size slope is parameterised as $-q_s-q_m$~so the $q_s$~parameter measures the need for a break in the slope of the size-number distribution.}  Bodies in the dotted line are included as impactors, but they are not evolved, but instead forced to obey a collisional equilibrium power law from the smallest evolving bin (see text).}
  \label{fig:examplesizedis}
\end{figure}

Within each semimajor axis bin there are 90 mass bins covering masses from $10^{10}$~to $10^{25}$~grams, initially equally spaced in log(mass).  The bin masses are not exactly fixed, but allowed to float to improve the resolution \citep[as in, e.g.,][and our implementation is the one tested in detail in \citet{2015ApJ...801...15S}]{1998AJ....115.2136K}.  To keep computation times down, but accurately model the collisional evolution of the smaller bins and avoid edge effects, bodies from $10^{-12}$~to $10^{10}$~grams are represented with non-evolving bins, whose evolution is not calculated, and whose size-number distribution is forced to obey $dn/ds \propto s^{-3.5}$~\citep{1969JGR....74.2531D}, with the normalisation fixed by the lowest mass evolving bin \citep[somewhat analogous to the procedure used in, e.g.,][]{2008ApJS..179..451K}.  Within the evolving bins the initial size distribution is a doubly broken power law, with the large size slope $q_L$, and the large break size $s_L$ set to their observed values, and the small break size $s_S$~set by inference from the cratering record of Charon, while the medium size slope $q_m$, and the small size slope $q_s$~are free parameters we introduce to fit with the model.  \citet{2020PSJ.....1...40K} explore producing $s_L$~and $s_S$~analogues by collisional evolution, and find that producing these breaks in the Cold Classicals would require a large total mass that is unlikely to be plausible for that population, so we take it to be the case that these breaks may be primordial.  For the Hot Classicals, the Scattering, and the Resonant populations; the large size slope is set to $q_L = -5.5$~\citep{2016AJ....152...23V}, and the large break size is set to $s_L = 55~\rm{km}$ \citep{2014ApJ...782..100F}.  Setting these three populations to the same values is partially motivated by the idea they formed from a single parent population that originated closer to the Sun \citep{2008Icar..196..258L}.  For the Cold Classicals, the large size slope is set to $q_L = -8.5$, and the large size break set to $s_L = 70~\rm{km}$~for $a < 50~\rm{au}$~\citep{2014ApJ...782..100F}, and within our model we choose to allow the large break size to scale with semimajor axis as:
\begin{equation}
    s_L = 70~\rm{km} \left(\frac{40~\rm{au}}{a}\right)^{\gamma}
    \label{eq:sl}
\end{equation}
where $\gamma$~is a free parameter we introduce to characterise our supposition that the characteristic size of formation may vary with orbital distance.  We will fit this $\gamma$~to the data.  The resulting initial size-number distribution that is used in our model is sketched in figure \ref{fig:examplesizedis}.

To model the evolving dynamical history of the Solar system, we deplete the unstable populations exponentially, with timescales of $1~\rm{Gyrs}$~for the Scattered disk \citep{1997Sci...276.1670D}, $3.43~\rm{Gyrs}$~for the 3:2 resonant objects, and $2.37~\rm{Gyrs}$~for the 2:1 resonant objects \citep{2009AJ....138..827T}.

To set the normalisation in each bin, we first assume the surface density of objects varies as $a^{-1.5}$~\citep{1977Ap&SS..51..153W}, then set the total number of large objects to match the present day populations from \citet{2014ApJ...782..100F} for the Cold Classicals and Hot Classicals, and the Outer Solar System Origins Survey (OSSOS) from \citet{2016AJ....152...23V}~for the Resonant populations, and from \citet{2018AJ....155..197L} for the Scattering population, accounting for the dynamical depletion imposed and assuming that collisional deletion for objects at $\gtrsim 50~\rm{km}$~will be negligible.  The latter assumption is predicted by theory \citep{2005Icar..173..342P, 2012A&A...540A..30V}, and now supported by various lines of observational evidence \citep{2012ApJ...744..139P, 2020Sci...367.6620M}.  In the case of the Cold Classicals, the observational normalisation is set for only the first bin, from $42~\rm{au} - 52~\rm{au}$, where the Main Cold Classical belt is located and cover the known objects.  Beyond that we posit there may be an extension of the Cold Classical belt to much larger semimajor axes.  We want to allow for the possibility that the primordial surface density may have discontinuously changed with radial distance, so we multiply surface density by a factor $\Pi$, which we introduce to quantify that change, and which we will fit to the data.  If the size-number distribution is constant as a function of semimajor axis, then $\Pi$~must be $\ll1$~to be consistent with the small number of large $\left( \gtrsim 100~\rm{km} \right)$ bodies detected by surveys in the optical \citep{2016AJ....152...70B}.  However, because we also let the large break size $\left(s_L\right)$~be a function of semimajor axis, $\Pi$~can be greater than one, if the characteristic size of large bodies decreases with semimajor axis (i.e., if $\gamma > 0$, equation \ref{eq:sl}), so we only restrict its value to be $\Pi \geq 0$.  We motivate $\Pi > 1$~cases partially by the argument the sweeping of the 2:1 MMR may have depleted the original population interior to its present location \citep{2015AJ....150...68N, 2017NatAs...1E..88F}, and partly by the observation that the surface density in solids is not a monotonically decreasing function of distance in protoplanetary disks \citep{2015ApJ...808L...3A,2018ApJ...869L..41A}.

The number of bodies in all bins evolve by mutual collisions.  The bodies in one semimajor axis and mass bin collide with those in all other bins with a frequency of $n \sigma v$, where $n$~is the volume density of those bodies, $\sigma$~is the cross section for a collision, given by $\pi\left(s_1 + s_2\right)^{2}$, where $s_1$~and $s_2$~are the radii of the two objects, and $v$~is the random velocity between the two bins; given by \citep{1993Icar..106..190W}
\begin{equation}
    v^2 = v_{1,kepler}^2 \times \left(\frac{5}{8}e_1^2 + \frac{1}{2}i_1^2\right) + v_{2,kepler}^2 \times \left(\frac{5}{8}e_2^2 + \frac{1}{2}i_2^2\right)
\end{equation}
where $e_1$~and $e_2$~are the eccentricities of the two bins; $i_1$~and $i_2$~are the inclinations of the two bins, and $v_{1,kepler}$~and $v_{1,kepler}$~are the Keplerian velocities at the middle of the two bins.  When two bins partially overlap, the interaction rate is decreased by the fractional overlap of the two bins, as outlined in \citet{2005Icar..174..105K}.  Collisional outcomes are determined using the prescription of \citet{2012ApJ...745...79L} for collisional outcomes, including disruption, super-catastrophic disruption, and hit-and-run collisions; our choice of model parameters prevent merging collisions.  We slightly simplify the calculations by always setting the impact angle to the median value of $60\degree$.  {\revised We use the material parameters from \citet{2009ApJ...691L.133S} fit to weak low density rock or sand with their parameters $\mu = 0.4$, $\phi = 7$, $q_s = 500$~(their $q_s$, different from our $q_s$), $q_g = 10^{-4}$, and a material density $\rho = 1 $g/cm$^3$.}

To explore what combinations of parameters can produce the presently observed Solar system, we employ the Markov-chain Monte Carlo code of \citet{2013PASP..125..306F}.  The model is allowed to explore four free parameters: the medium size slope $q_m$, and the small size slope change $q_s$, which determine the primordial size-number distribution of the bodies in all populations, $\gamma$, a power-law exponent that scales the large break size of the primordial size number distribution of Cold Classical objects with formation distance, and $\Pi$, a scaling factor for the surface density of Cold Classical bodies beyond 50 au.  The goodness of fit for each trial is determined by comparisons to these observations: 
\begin{itemize}
    \item The cratering record of Charon as reported in \citet{2019Sci...363..955S}; we parameterise their Relative (R)-plot of the surface density of craters as $R\left(s = 100~\rm{m}\right) = 0.004 \pm 0.002$, $R\left(s = 500~\rm{m}\right) = 0.05 \pm 0.025$, and $R\left(s = 7.5~\rm{km}\right) = 0.05 \pm 0.025$, where these values are approximately estimated from the dispersion within that work.  The volume density of TNOs in space is converted to R using the impact probabilities for each population onto Charon calculated by \citet{2015Icar..258..267G};
    \item the on sky surface density of occulting bodies with $r > 250~\rm{m}$ in the ecliptic today from the observations of \citet{2012ApJ...761..150S}, who found $N(r > 250~\rm{m})=1.1_{-0.7}^{+1.5} \times 10^7 \ \mathrm{deg}^{-2}$;
    \item the on sky surface density of occulting bodies with $r > 530~\rm{m}$ in the ecliptic today compared to the observations of \citet{2020submitted}, who found $N(r > 530~\rm{m})=1.0_{-0.4}^{+0.2} \times 10^7 \ \mathrm{deg}^{-2}$;
    \item the pencil-beam optical survey of \citet{2004AJ....128.1364B}, who surveyed $0.019$~square degrees in the ecliptic, with a modelled detection efficiency of $\sim 100\%$~for $m < 28.4$, and $\sim 50\%$ for $28.4 < m < 28.6$.  They discovered $3$~objects in the classical populations, which we assign an uncertainty of $\sqrt{3} \approx 1.7$~objects \citep{2006AJ....131.2364B}.
\end{itemize}

\begin{figure}
  \includegraphics[width=\linewidth, trim = 0 0 0 0]{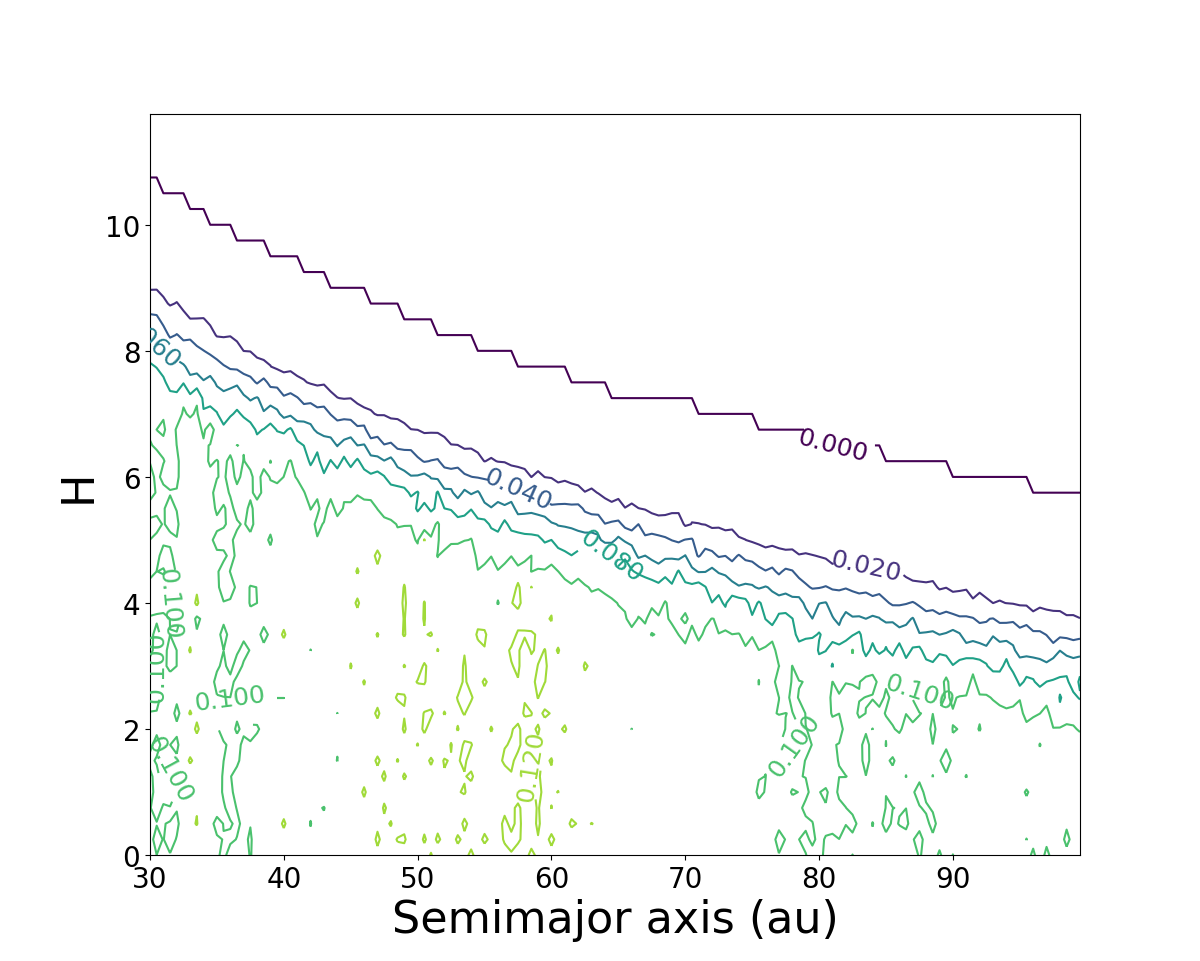}
  \caption{The fraction of bodies in a cold disk detected by the OSSOS survey simulator, calculated using the method outlined in \citet{2018FrASS...5...14L}, as a function of absolute magnitude $H$~and semimajor axis $a$.  Each point considered 1 million hypothetical objects, so the 0.000 line represents a less than 0.0001\% chance of discovery.}
  \label{fig:OSSOS}
\end{figure}

{\revised Because the direct observations and occultations measure bodies that are in the ecliptic, all populations are measured in it, with more highly inclined populations contributing a smaller fraction of their bodies to those measures.  Cratering requires the orbits of a population to overlap that of Pluto-Charon, all populations except the Cold Classicals beyond $50$ au contribute to the cratering, with the 3:2 resonance the most highly overlapping.}
There exist other possible observational tests we choose not to employ.  We use the OSSOS results to set the input, rather than evaluate the output, to keep the number of free parameters reasonably low, and as we are primarily interested in smaller $\left(\sim 1~\rm{km}\right)$~bodies, which are not well measured by that survey.  In figure \ref{fig:OSSOS}, we use the OSSOS simulator \citep{2018FrASS...5...14L} to understand what Cold Classical objects might be detected by optical surveys.  For this calculation we use a uniform flat distribution of eccentricity from $e = 0$~to $e = 0.04$.  The default model employed by OSSOS has eccentricity increase with semimajor axis \citep{2011AJ....142..131P}, but whether this should expected to continue to larger semimajor axis is unclear \citep[e.g., see][]{2011ApJ...738...13B}, so instead we consider a cold population that remains dynamically cold at all semimajor axes.  Exact conversion of absolute magnitude to sizes depends on the albedo of the bodies, which is known to have varied with the semimajor axis of formation \citep{2014ApJ...782..100F}, as is not well observationally constrained beyond $\sim 46~\rm{au}$, but for reference comets have a typical albedo of $0.04$, small Hot Classicals of $0.06$, and Cold Classicals of $0.15$, which for a spherical body of $100~\rm{km}$~diameter corresponds to $H \approx 9.1, 8.7$~and $7.7$. 
Although the 2:1 mean-motion resonance with Neptune is sometimes associated with the outer boundary of the Cold Classical Kuiper belt, \citet{2018ApJS..236...18B} identifies four apparent detections of Cold Classical bodies beyond the 2:1 mean-motion resonance at 47.8 au; they identified those bodies by selecting those objects with $q > 40~\rm{au}$~and inclinations of at most a few degrees.  The exact choice for cutoff parameters can slightly affect the count, for instance, their object $o5d052$, with $a = 53.7~\rm{au}$, $i = 4.5 \degree$, but $e = 0.29, q = 38.2~\rm{au}$~is classified by their criteria as a detached object, but might plausibly be better associated with the Cold Classicals.  Although these are only a handful of objects, the outer limits of the Cold Classical belt are not well established, and limits depend on the assumptions made about the properties of the bodies in the population.

We convert the bodies from the model populations into an expected number of objects within $\beta$ of the ecliptic following \citet{2014ApJ...782..100F}:
\begin{equation}
    F(<\beta) = \frac{2}{\pi} \sin^{-1}{\left(min\left[\frac{\sin{\beta}}{\sin{i}}, 1\right]\right)}
\end{equation}
which allows us to calculate what fraction of the time a body with inclination $i$~is within $\beta$ of the ecliptic, for comparison to our observational tests.

\subsection{Ensemble Model Evaluation}

We initialise our default case with 16 walkers searching the phase space, initially dispersed around the point $\Pi = 1$, $q_m = 3$, $q_s = 4$, and $\gamma = 4$, values chosen to fit observational constraints and found to produce plausible model fits in initial evaluations.  In some other cases, parameter choices can result in models that are both bad fits to the observational tests and extremely computationally onerous to evaluate.  These can prevent initialisations from converging in a reasonable computation time, but are not a significant problem later on, as later in searchs they are found to be much worse fits, and thus the walkers typically revert to their previous step location.  To randomise the walkers, we begin the search with the 16 walkers spread around the starting point by perturbing the initial values randomly by $0.1$.  The searches were constrained to the space defined by $0 \leq \Pi \leq 100$, $0.0 \leq q_m \leq 6.0$, $-10.0 \leq q_s \leq 10$, and $-5 \leq \gamma \leq 15$.

\begin{figure}
  \includegraphics[width=\linewidth, trim = 0 0 0 0]{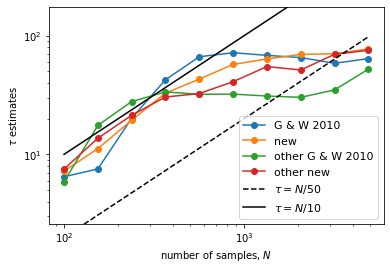}
  \caption{Autocorrelation time estimates for two different sets of MCMC chains for the $q_m$~parameter, estimated in two different ways, using the method of \citet{2010GW} and of \citet{2013PASP..125..306F}.  The four estimates are slightly different, but all indicate a correlation time of $\lesssim 10^2$.  The other parameters also have autocorrelation times of $\mathcal{O}\left(\tau\right) \sim 10^2$, with $\tau_\gamma$~a factor of $2\sim3$~larger.}
  \label{fig:qm_chains}
\end{figure}

\begin{figure}
  \includegraphics[width=\linewidth, trim = 0 0 0 0]{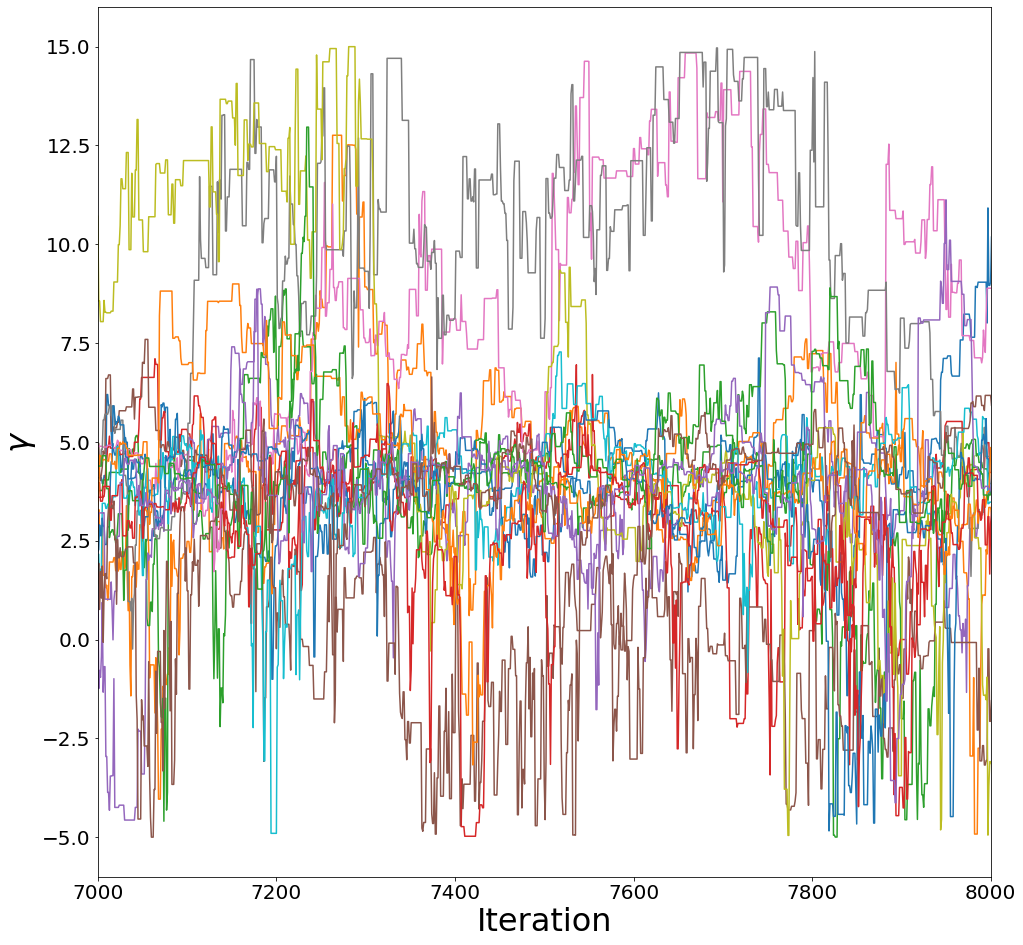}
  \caption{MCMC Chains for the model parameter $\gamma$, after the chains are well converged.  The walkers mostly stay near the median value of $3.6$, but some of the excursions to higher or lower values are somewhat 'sticky', reflecting local minima in the probability space.}
  \label{fig:gamma_chains}
\end{figure}

Examination of the autocorrelation times (Figure \ref{fig:qm_chains}) in the Markov chains using the approaches suggested in \citet{2013PASP..125..306F}, suggest a typical value of $\mathcal{O}\left(10^2\right)$.  Because the model is multidimensional and nonlinear, there are local minimums in the parameter space where the walkers can become stuck for more extended periods (Figure \ref{fig:gamma_chains}).  For our model evaluation, we run the chains for $13500$~steps and discard the first $9000$.  Because our chains are significantly longer than the autocorrelation times, the chains will be well converged.

\begin{figure}
    \centering
    \includegraphics[width=\linewidth, trim = 0 0 0 0]{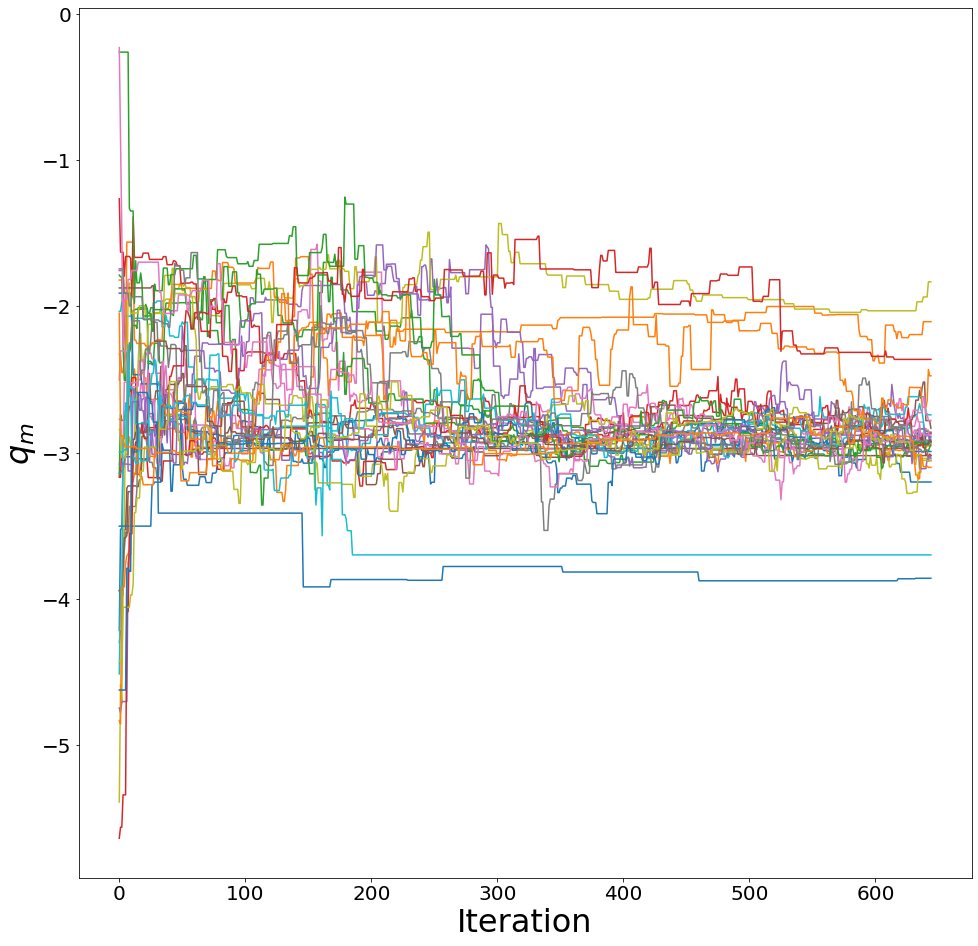}
    \caption{In this example, we initialise 32 walkers randomly and uniformly across the full allowed space $0 \leq \Pi \leq 100$, $0.0 \leq q_m \leq 6.0$, $-10.0 \leq q_s \leq 10$, and $-5 \leq \gamma \leq 15$.  The walkers are converging towards the same $q_m \sim 2.9$~than the longer run, but the model evaluation times for some points far from the good solutions can be exceedingly long, making it computationally difficult to fully evaluate such chains.  Because of this, we prefer to start walkers near our default solution.}
    \label{fig:qm_chains_full}
\end{figure}

Our starting point for the MCMC walkers is physically and observationally motivated, but we wish to be confident it does not represent some kind of local minimum within the parameter space.  To test this, we initialise a run of 32 walkers randomly and uniformly across the allowed search space ($0 \leq \Pi \leq 100$, $0.0 \leq q_m \leq 6.0$, $-10.0 \leq q_s \leq 10$, and $-5 \leq \gamma \leq 15$).  Initial computations are slow, but within $\sim 10^2$~iterations roughly half the walkers are within the converged solution space (Figure \ref{fig:qm_chains_full}).  The remaining walkers continue to converge towards the best fit solutions, but the convergence slows and the increased computation times at some of the "sticky" points prevent us from running this case to complete convergence.  Nonetheless, the results here give us confidence that our choice of initial conditions produce good coverage of the applicable solution phase space.

\section{Results}

\label{sec:results}

\begin{figure*}
   \includegraphics[width=\linewidth, trim = 10 0 0
   0]{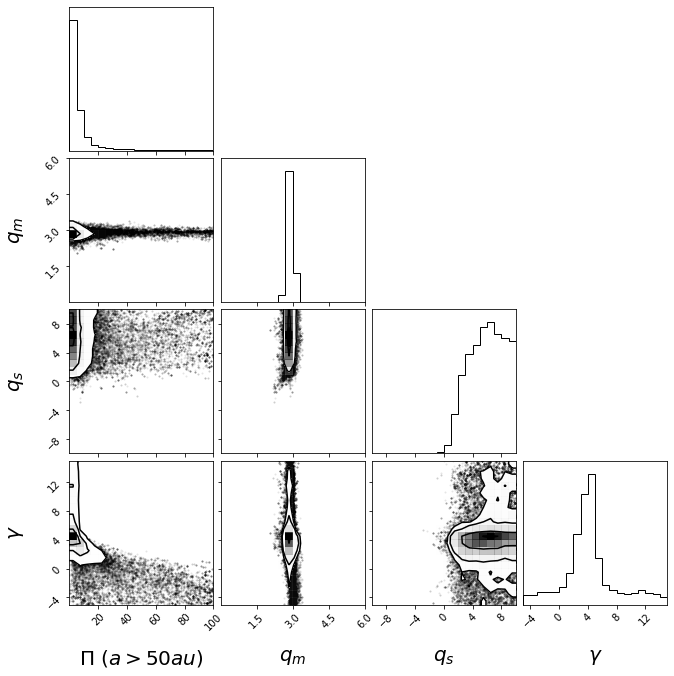}
  \caption{A corner plot of the results of the MCMC search of the viable parameter space.  The contour plots show the correlations between parameters, the histograms the one parameter distribution within the model evaluation.  The best fit values of the model parameters are $\Pi = 4.4^{+16.1}_{-2.4}$, $q_m = 2.9^{+0.1}_{-0.1}$, $q_s = 6.3^{+2.4}_{-2.9}$, and $\gamma = 3.6^{+1.4}_{-3.1}$, where $q_s$~can probably be increased to arbitrarily large values and remain a good fit.  $q_m$ is strongly constrained, while the other parameters have significant areas of lower fit quality that are not strongly excluded.}
  \label{fig:corner_0}
\end{figure*}

The results of the full chains are plotted in figure \ref{fig:corner_0}.  The best fit values of the model parameters are $\Pi = 4.4^{+16.1}_{-2.4}$, $q_m = 2.9^{+0.1}_{-0.1}$, $q_s = 6.3^{+2.4}_{-2.9}$, and $\gamma = 3.6^{+1.4}_{-3.1}$.  The best fit $\Pi$~does not indicate the model has a preference for a jump in the surface density, as examination of the distribution of values shows they are clustered near $1$, with a long tail to large values, whereas negative values were disallowed as unphysical.  The lower bounds on $q_s$~are produced by the model, but the upper bounds appear to be produced by the choice to not allow values of $q_s > 10$.  We do not reconsider this limit, as we expect and confirm that very large values quickly converge in behaviour by collisional evolution.  The four free parameters do not show significant correlations, except in the region of large $\Pi$~and small/negative $\gamma$ (figure \ref{fig:corner_0}, bottom left panel), which are disfavoured but not entirely excluded by the model.  In these cases, the total mass increases with distance, but that mass is largely in a small number of very large bodies, resulting in a poor but not completely excluded fits by the model.  It might be possible to further exclude this area with the addition of further observational tests sensitive to small numbers of large objects (such as shallow all sky surveys like \citet{2015AJ....149...69B}), but as these represent similar models at the {\revised roughly} $100~\rm{m}$~{\revised to} $\sim 1~\rm{km}$~sizes we are interested in, we chose not to pursue such complications.

To better understand the solution, we plot an example of the evolution for {\revised the best fit} parameters in figure \ref{fig:example_colds}.  From it, we can see how the model meets the various observational criteria.  The size of the largest bodies that formed decreases with distance from the Sun, keeping the number of large bodies that could be seen in optical surveys small.  Even though the total mass in each radial bin is decreasing with distance from the Sun (apart from a small bump after the first bin) the number of bodies at a few hundred meter sizes increases, as the mass is distributed there, rather than in larger bodies, producing large numbers of bodies a few hundred meters in size that can be detected in serendipitous occultation surveys.  These bodies {\revised farther from the Sun} do not produce a cratering record on Pluto or Charon, as their orbits do not overlap.  In this way, the parameters produced from the model can be tied to specific observables, for instance the $\Pi$~parameter only affects bodies that do not produce a cratering record on Pluto or Charon, so that observational test does not favour any particular value for $\Pi$~(Figure \ref{fig:upfactorversuscraters}).  In contrast, the pencil-beam HST survey of \citet{2004AJ....128.1364B}, disfavours large values of $\Pi$ (Figure \ref{fig:pivscraters}), as that survey is sensitive to bodies with radii as small as 10 km at 50 au.  {\revised The bodies below $\sim 100~m$~do not appear in any of the observations, so the effective slope measured for the smallest bodies is $\sim -2$.  If smaller bodies were measurable, we should expect them to have a more negative slope as our fit needs for a positive $q_s$, which is driven primarily by the observation of craters at 200 m. However, as with all parameters there is a complex interaction between the four input parameters and six observations. We can also compare the simulated size distribution of bodies in the main Kuiper belt (dark blue line in Figure~\ref{fig:example_colds}) to that obtained from Charon's cratering record (light blue curve in Figure~\ref{fig:smallBodyObservations}) for the smallest sizes, i.e., between radii of 100 m and 500 m. We notice that the simulations show a dip at $\sim$ 500 m before transitioning to the -3.5 slope at smaller sizes. This dip may explain the shallower slope in -1.7 observed between 100 and 500 m from the Charon's cratering record. However, the dip does not extend to a wide range of radii and seems to not fully explain the observations. The simulations show the end state of the collisional evolution but it is expected that the dip was much deeper in the Solar system's youth (see dashed line in Figure~\ref{fig:example_colds}), when most impacts happened with Charon. Because the cratering record with Charon is integrated over the whole history of the Solar system and the simulation is at a given time, we should not directly compare the two figures but it seems that indeed the dip (that was even deeper early on when most impact happened) goes in the right direction to explain this observed shallower slope in -1.7.

\begin{figure}
  \includegraphics[width=\linewidth, trim = 0 0 0 0]{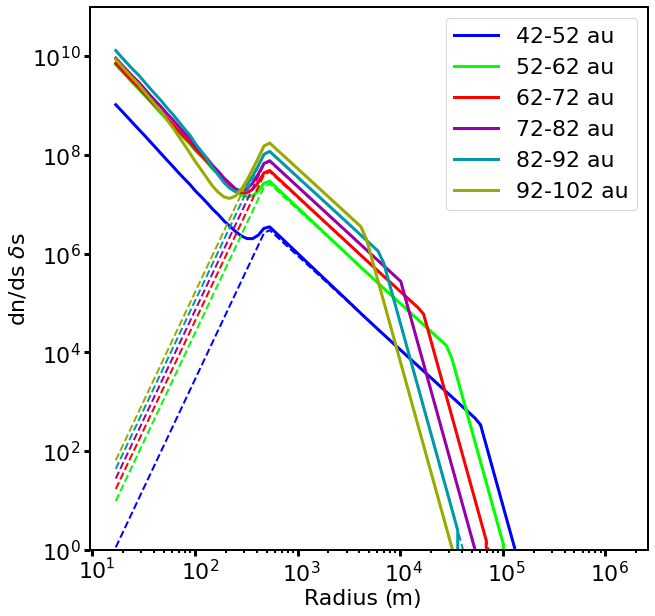}
  \caption{Initial (dashed lines) and final (solid lines) size-number distributions of the Cold bodies, from a simulation that begins with $\Pi = 4.4$, $q_m = 2.9$, $q_s = 6.3$, and $\gamma = 3.6$.  The simulation also includes Hot, Resonant, and Scattering populations, which are not plotted here for clarity, but for which the overall evolution is similar.  {\revised Collisional Evolution is largely confined to the sizes below $\sim 500 m$.  The number of small bodies increases in farther out bins even as the total mass declines as the $\gamma$~parameter pushes the mass distribution towards smaller sizes.  The $\Pi$~parameter creates the larger discontinuity between the first and second radial bins.}}
  \label{fig:example_colds}
\end{figure}

\begin{figure}
  \includegraphics[width=\linewidth, trim = 0 0 0 0]{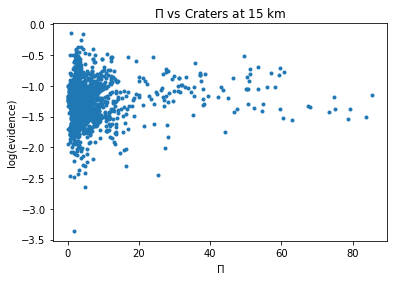}
  \caption{Log probability contributed by the cratering record at 15km from various models runs, plotted against the upfactor $\Pi$.  The scatter is induced as these are the results from calculations from the full model, and thus also span a range of $q_m$, $q_s$, and $\gamma$.  Because $\Pi$~dictates the size-number distribution at larger semimajor axis, it has a negligible influence on the cratering record of Charon.}
  \label{fig:upfactorversuscraters}
\end{figure}

\begin{figure}
  \includegraphics[width=\linewidth, trim = 0 0 0 0]{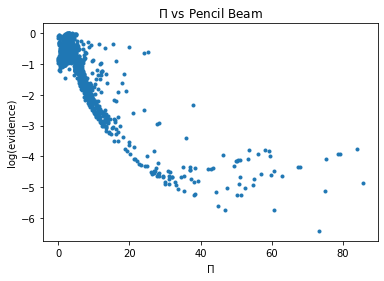}
  \caption{Log probability contributed by the pencil beam survey of \citet{2004AJ....128.1364B} from various models runs, plotted against the upfactor $\Pi$.  The pencil beam survey disfavours high values of $\Pi$, as large numbers of $s \sim 10~\rm{km}$~bodies at 50+ au would be detectable in that survey.}
  \label{fig:pivscraters}
\end{figure}

In the example evolution, the initial conditions are fairly similar to the present day conditions, {\revised and in this case} significant collisional evolution only occurs at sizes of a few hundred meters and smaller, {\revised where the more numerous, smaller bodies collide more frequently.  With  steep size-number distribution, the transition from where the chance of a catastrophic collision is $ \gg 1$~to where it is $ \ll 1$~is quite sharp}.  This low amount of collisional evolution is consistent with the observation that (486958) Arrokoth is not significantly collisionally evolved \citep{2020Sci...367.3999S}, nor are the ultra-wide binaries of the Cold Classicals \citep{2012ApJ...744..139P}.  

We noted that examination of the distributions suggests our choice to restrict $q_s \leq 10.0$~is likely excluding larger values of $q_s$~that would give acceptable fits.  In this example evolution, we can see that because {\revised $q_s$}~governs the size distribution below $500~\rm{m}$, which is significantly filled in simulations that begin with very few small bodies, it is likely there is no upper limit for this parameter.  If $q_s + q_m \gtrsim 0$, smaller bodies are less numerous than larger bodies (figure \ref{fig:examplesizedis}), and thus the larger bodies drive the collisional evolution, which results in all such models quickly converging.  Thus we are safe not considering larger values of $q_s$~here.

\section{Discussion}

\label{sec:discussion}
\subsection{Might formation size vary with formation distance?}

In the Streaming Instability model, simulations find that the formation size of planetesimals will vary with distance \citep{2016ApJ...822...55S}.  In a simple Minimum Mass Solar Nebula model \citep{1977Ap&SS..51..153W}~this size increases slowly with formation distance, but such a simple model may not reflect reality \citep{2007ApJ...671..878D,2017ApJ...839...16C,2020ARA&A..58..483A}, so this remains a possibility.  \citet{2020ApJ...901...54K} find that for their assumptions about disk evolution, the formation size slowly increases with distance, but turns over and quickly decreases at 10s of au.  \citet{2021MNRAS.500..718K} suggested that large populations of small $\left( \sim~\rm{km} \right)$~without accompanying large $\left(\sim 100~\rm{km}\right)$~bodies might be able to resolve some issues with debris disk models that otherwise infer implausibly large masses.  If this interpretation is correct, it would be natural to expect a similar population could exist around the Sun.

\subsection{Can we find small, distant bodies?}

Fundamentally, the existence of a model that can fit all the observations is not a demonstration that the model represents the astrophysical reality.  This concern is perhaps particularly strong in a case like this where the model has tunable parameters that give it significant flexibility.  This naturally poses the question of how we can demonstrate the detections are of astrophysical occulting objects, rather than some kind of noise of the instruments or artifacts of the data reduction.

Occulting objects are far too small to be followed up (with expected apparent magnitudes $\gtrsim 35$), even by the most optimistic cases for LUVOIR \citep{2019arXiv191206219T}.  The most straightforward evidence would be to demonstrate that the occultations preferentially occur in the ecliptic.  Previous occultation surveys have been limited in their ability to do this.  \citet{2012ApJ...761..150S} reported two detections using the Hubble Space Telescope's fine guiding sensors.  They reported detections at $6.6\degree$, $14.4 \degree$, and a possible detection at $81.5 \degree$, from a sample that was roughly uniform across the sky.  These detections are concentrated towards the ecliptic, but that result is not statistically significant.  \citet{2019NatAs...3..301A} reported the detection of a single object at $0.2 \degree$, but imaged only stars within $0.8 \degree$~of the ecliptic, and thus the detection has insufficient statistical power to demonstrate the detections are concentrated towards the ecliptic.  \citet{2015MNRAS.446..932L} reported 13 possible detections in \textit{CoRoT}~fields around $20 \pm 8 \degree$~from the ecliptic, again limiting their ability to measure the inclination distribution of the potential detections.  \citet{2020submitted} reported five detections in fields from $0 \degree$~to $90 \degree$, but the searched fields are concentrated towards the ecliptic, and thus the detections do not have the statistical power to convincingly demonstrate that they are not uniform across the sky.

The large TNOs generally follow an inclination of the form \citet{2001AJ....121.2804B} :
\begin{equation}
    P\left(i\right) \propto sin\left(i\right) e^{-\frac{1}{2}\left(\frac{i}{\sigma}\right)^2}
\end{equation}
with two components with $\sigma \approx 16\degree$~and $\sigma \approx 2.6\degree$, with the former component about twenty times more massive than the latter \citep{2014ApJ...782..100F}.  However, this does not perfectly fit the distribution, and a small number of bodies can be found at very high inclinations, due to various mechanisms \citep{2011Icar..215..661G,2016ApJ...827L..24C}, so a single detection at very high inclination is not sufficient to demonstrate non-astrophysicalness.  Nonetheless, we can estimate the number of objects we would need to detect to satisfy ourselves the potential events are indeed occultations by objects within the Solar system.  In figure \ref{fig:NeededOccultations}, we plot the Kolmogorov-Smirnov statistics of a Monte Carlo suite of realisations of MIOSOTYS surveys that detect $N$~real TNOs, compared to non-physical detections randomly and uniformly distributed across the survey area.  For each $N$, we realise 100 surveys, all using the same search areas on the sky as the MIOSOTYS survey presented in \citet{2020submitted}, to understand the distribution.  In that plot, we assume that the $\sigma \approx 16\degree$~and $\sigma \approx 2.6\degree$~components have equal numbers of bodies at sizes of a few hundred meters, reflecting the model presented in this work.  Examining the figure, we can see that we would typically expect to need $15-20$~ detections to be 99\% confident the distribution is not uniform on the sky.  In the case where the Hot population outnumbers the Cold population by a factor of $\sim 20$, 4-5 times as many detections are needed to exclude a uniform distribution to the same confidence.  Although we use the MIOSOTYS footprint, the details of the survey implementation are not too important; a survey that spent half its time in the ecliptic and half its time on the ecliptic pole would only need half as many detections to obtain the same level of confidence the signal was astrophysical, but would need twice as much survey time to obtain the detections, as only the observations in the ecliptic would produce detections.

\begin{figure}
  \includegraphics[width=\linewidth, trim = 0 0 0 0]{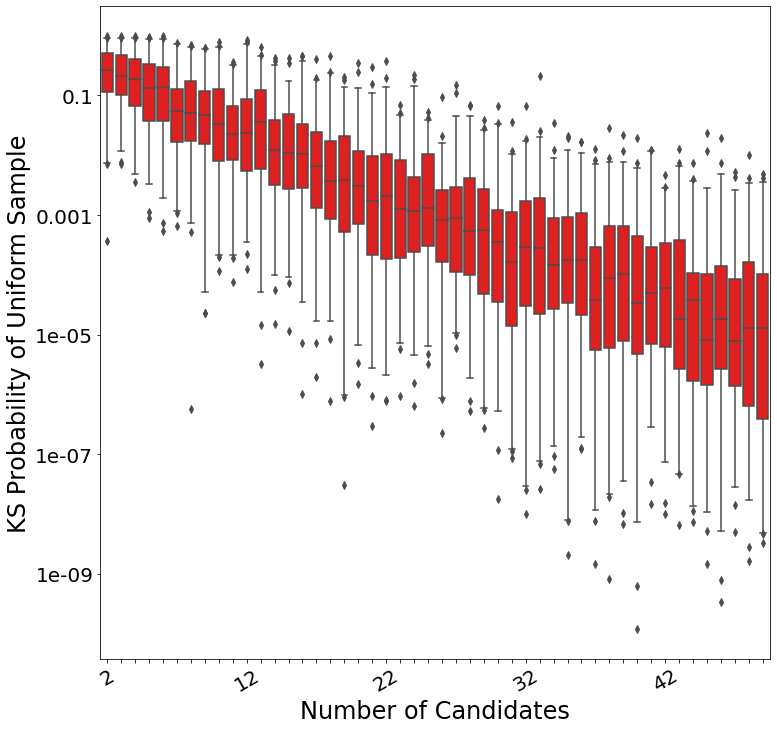}
  \caption{A Box-and-Whiskers plot of the distribution of the Kolmogorov-Smirnov statistics comparing astrophysical detections to sky-uniform detections, with 100 realisations of surveys of the MIOSOTYS survey area that report $N$~candidate detections, for $N$~from 2 to 50.  The red boxes contain 50\% of the data points, and the whiskers 96\%, which the other trials plotted as outlier points.}
  \label{fig:NeededOccultations}
\end{figure}

For future proposed missions to the outer Solar system  \citep[e.g.,][]{2019AcAau.162..284M,2020A&A...641A..45H,2021BAAS...53d.039T}, it could be possible to detect an extended cold belt.  If the mission is close to the ecliptic, the brightest member would have an apparent magnitude of $\sim 13$~in the best fit model.  \emph{New Horizons} detected TNOs to $m \sim 19$~\citep{2019AJ....158..123V}, so such aspirations are plausible.  However, any space mission is virtually assured to be far more costly than an extended MIOSOTYS or similar programme, so the latter will remain the plausible approach.

{\revised
\subsection{Can we plot all the observations together?}
So, then how can we understand the apparent contradiction of figure \ref{fig:smallBodyObservations}?  When we put the directly observed, cratering, and occulting populations on the same plot, we implicitly (or explicitly) think of them as coming from the same population.  But the radial sensitivity of the three observations are very different, and perhaps it is wiser to start from a place of skepticism towards them representing the same population.  Certainly, if the model presented here is correct, the observations should not be presented side by side.
}

\section{Conclusions}

\label{sec:conclusions}

We consider models where the Cold Classical Kuiper belt extends to larger semimajor axes than the known objects, and the size at which the bodies form is allowed to change with their formation distance.  We find that models where the formation size decreases with increasing semimajor axis can produce both the relatively high number of few hundred meter sized bodies seen in occultation surveys and the relatively low number inferred from the cratering record on Pluto/Charon.  The most promising method of testing the astrophysical reality of these models is extend occultation surveys with broad inclination coverage to obtain sufficient events to measure their inclination distribution across the sky.

\section*{Acknowledgements}

We wish to thank the anonymous referee for a thoughtful review that improved the paper. We thank Philippe Th\'{e}bault for useful discussions.  This work received funding from the European Research Council under the European Community’s H2020 2014-2020 ERC Grant Agreement No. 669416 ``\emph{Lucky Star}".

{\footnotesize
\bibliographystyle{aa}
\bibliography{FinalManuscript}
}

\end{document}